\documentclass[preprint,tightenlines,11pt,superscriptaddress,nofootinbib,pra,preprintnumbers,amsmath,amssymm,
twocolumn,
superscriptaddress,floatfix]{revtex4-1}

\pdfoutput=1
\usepackage[pdftex]{graphicx}
\usepackage{dcolumn}% Align table columns on decimal point
\usepackage{bm}% bold math
\usepackage{graphics}
\usepackage{rotating}
\usepackage{amssymb} 
\usepackage{comment}
\usepackage{multirow}
\usepackage{footnote}
\usepackage{comment}
\usepackage[colorlinks=true]{hyperref}
\usepackage{url}

\usepackage[dvipsnames]{xcolor}
\usepackage{tikz}
\usetikzlibrary{calc}% neded for coordinate calculations
\usetikzlibrary{arrows}
\newcommand{\midarrow}{\tikz \draw[-triangle 90] (0,0) -- +(.1,0);}
\newcommand{\Depth}{6}
\newcommand{\Height}{-6}
\newcommand{\Width}{6}
\usepackage{subfigure}

\DeclareMathOperator{\Tr}{Tr}

\renewcommand{\Re}{{\text{Re}}}

\newcommand{\fermilab}{Fermi National Accelerator Laboratory, Batavia, Illinois, 60510, USA}
\newcommand{\cuboulder}{Department of Physics, University of Colorado, Boulder, Colorado 80309, USA}
\newcommand{\rbrc}{RIKEN BNL Research Center, Brookhaven National Laboratory, Upton, New York 11973, USA}

\def\today{\number\day\space\ifcase\month\or
January\or February\or March\or April\or May\or June\or
July\or August\or September\or October\or November\or December\fi
\space\number\year}
\newcount\mins \newcount\hours
\def\now{\hours=\time \mins=\time
	\divide\hours by60 \multiply\hours by60 \advance\mins by-\hours
	\divide\hours by60 
	\number\hours:\ifnum\mins<10 0\fi\number\mins }

\begin{document}

\title{Digitizing Gauge Fields: Lattice Monte Carlo Results for\\ Future Quantum Computers}

\author{Daniel C. \surname{Hackett}}
\email[]{daniel.hackett@colorado.edu}
\affiliation{\cuboulder}

\author{Kiel \surname{Howe}}
\email[]{khowe@fnal.gov}
\affiliation{\fermilab}

\author{Ciaran \surname{Hughes}}
\email[]{chughes@fnal.gov}
\affiliation{\fermilab}

\author{{William~\surname{Jay}}}
\email[]{wjay@fnal.gov}
\affiliation{\cuboulder}\affiliation{\fermilab}

\author{Ethan~T.~\surname{Neil}}
\email[]{ethan.neil@colorado.edu}
\affiliation{\cuboulder}\affiliation{\rbrc}

\author{James~N.~\surname{Simone}}
\email[]{simone@fnal.gov}
\affiliation{\fermilab}

\preprint{FERMILAB-PUB-18-615-T}

\begin{abstract}

In the near-future noisy intermediate-scale quantum (NISQ) era of quantum computing technology, applications of quantum computing will be limited to calculations of very modest scales in terms of the number of qubits used.  The need to represent numeric quantities using limited resources leads to digitization errors which must be taken into account. As a first step towards quantum simulations of realistic high-energy physics problems, we explore classically the effects of digitizing elements of the $\mathrm{SU}(2)$ gauge group  to a finite set. We consider several methods for digitizing the group, finding the best performance from an action-preserving projection onto a mesh.  Working in (3+1) dimensions, we find that using $\sim 7$ (qu)bits to represent each $\mathrm{SU}(2)$ gauge link induces a digitization error on the order of $10\%$ in short-distance observables and $2\%$ in long-distance observables.  Promisingly, our results indicate that each $\mathrm{SU}(2)$ gauge link can be represented by $\mathcal{O}(10)$ (qu)bits, from which we estimate that a $16^3$ $\mathrm{SU}(2)$ lattice could be simulated with no more than $\mathcal{O}(10^5)$ (qu)bits.  Our results on digitization are also of interest as a form of lossy compression that could be used in high-performance classical computing to alleviate communications bottlenecks.

\end{abstract}

\maketitle

%===========================================================================8
% Section Seperator
%===========================================================================8

\section{Introduction}
\label{sec:intro}

Quantum computers offer the promise of solving problems which are presently intractable.  In particular, simulating strongly interacting gauge theories on a digital quantum computer is an exciting prospect.  Many interesting physics problems in lattice gauge theory remain intractable even for cutting-edge classical computers, including real-time dynamics of hadronization and thermodynamics at large quark-number density. While the lattice community has made impressive progress in studying hadronic physics directly from QCD \cite{Kronfeld:2012uk,Savage:2016egr,Beane:2010em}, the heavy nuclear physics needed by neutrino experiments, like the quark-current form factors of Argon, remain beyond the limits of current classical computers. More fundamentally, some simulations are intractable on \emph{any} classical computer. For example, storing the full wavefunction of a $500$-qubit system would require more classical bits than there are atoms in the observable Universe\footnote{On an $N$-bit classical computer, $2^N$ different numbers are accessible, but the memory needed to store $m$ numbers is $mN$-bits. On a $N$-qubit quantum computer, $2^N$ different numbers are also accessible but these $N$-qubits can also store the $2^N$ values. }.

Despite the eventual promise of quantum computing, speculation about near-term prospects for these devices in the ``noisy intermediate-scale quantum'' (NISQ) era suggests that qubit and gate resources will be severely limited \cite{2018arXiv180100862P}.
On the other hand, due to their local nature, field theory Hamiltonians require less connectivity to implement than general quantum computing algorithms and may in fact be among the first realistic problems tractable on large-scale quantum computing devices. Nevertheless, efficient use of available qubit resource will be important to simulate a gauge theory on any near-future quantum computer. This requires us to confront ``digitization error'', which we define as the error due to representing continuously-valued quantum fields in finite numerical precision.

Field theories are physical systems with infinitely many degrees of freedom, and their simulation on any digital computer requires finite approximations.  The standard approach restricts the fields to a discrete spacetime lattice in a finite volume, and the control and extrapolation of the resulting {finite lattice-spacing error} and {finite-volume error} is well-known in the literature on lattice field theory.  However, there is a third finite approximation which is usually not emphasized: the {quantum fields themselves} must be restricted to take on discrete numerical values, limited by floating-point precision.  In principle, this discretization yields a third approximation error, {digitization error.}  

Of course, the use of $64$-bit double-precision floating point numbers guarantees that such errors will be utterly negligible in most modern lattice calculations, although there can be concerns about reversibility of the hybrid Monte Carlo algorithm in global sums on large lattices \cite{Urbach:2017ate}.  We refer to the common use of 64-bit floating point numbers to represent gauge fields as \emph{ultrafine digitization}.

Even in the early days of computational lattice field theory, the availability of classical bits was sufficient to avoid large digitization errors compared to other error sources.  However, there was early interest in simulating with discrete subgroups of $\mathrm{SU}(N)$, which would allow the use of lookup tables to speed up the computations \cite{Creutz:1983ev}.  These studies found that the subgroup digitization created an artifact lattice phase at weak coupling. For the $120$-element largest point subgroup of $\mathrm{SU}(2)$, this artifact phase was at sufficiently weak coupling that some parameters of interest remained accessible \cite{Rebbi:1979sg,Petcher:1980cq}. This was not the case for the largest $1080$-element point subgroup of $\mathrm{SU}(3)$ \cite{Bhanot:1981xp}. However, an additional study \cite{Lisboa:1982jj} found that this phase could be pushed to weaker coupling in $\mathrm{SU}(3)$ by using a finer digitization generated by interpolating between $\mathrm{SU}(3)$ elements.

In this work, our goal is different: we study the impact of extreme memory restrictions, rather than to improve computational efficiency.  We ask how coarsely the digitization of group elements onto a finite set may be done before inducing large systematic errors.  By translating the size of the digitized group into the number of (qu)bits required to represent a gauge link, we can estimate the number of (qu)bits required to avoid digitization error at some level.  We study several different group-truncated digitization schemes to gain insight into the features of an optimal scheme.

More specifically, we consider the effect of digitizing pure $\mathrm{SU}(2)$ gauge theory.
Although not directly relevant to QCD, this theory is computationally less expensive than $\mathrm{SU}(3)$, allowing us to generate large data sets and thereby avoid statistical errors that might introduce ambiguity in our results otherwise.
Furthermore, $\mathrm{SU}(2)$ is isomorphic to the four-dimensional unit sphere $S_3$, providing an intuitive picture for digitization that would be lacking for larger gauge groups.  Our work shares some characteristics with a recent paper by Urbach \cite{Urbach:2017ate}, although his focus is on the HMC algorithm while we focus on the observables, and his study is concerned with a much finer digitization than our results will explore.

Other work on gauge group digitization in quantum computing for high-energy physics has focused on finding few qubit problems that can be implemented on NISQ technology. For example, \cite{Klco:2018kyo} examines the $(1+1)$-D Schwinger model with one and two spatial sites using a $2$-qubit digital quantum computer and shows that the quantum computer produces observables which agree with analytic results for a large time range before fidelity becomes appreciable. Also, using an analogue quantum computer, \cite{Lu:2018pjk} computes two-body and three-body forces between heavy-mesons in the Schwinger model. The authors of \cite{Zohar:2013zla} formulate the $\mathrm{SU}(N)$ Hamiltonian with matter fields for an analogue quantum simulation and study the digitization error in the ($2+1$)-D compact QED ground state energy for a range of couplings. All these works digitize the gauge group by imposing a cutoff on the allowed energy eigenstates.  This approach has the advantage of conceptual clarity: keeping only the low-energy eigenstates should leave low-energy physical observables unaffected. However, it is unclear how the resource requirements of the eigenstate-truncation method compares with the group-truncation schemes that we study here, especially for short-range observables.

The outline of this work is as follows: in Section~\ref{sec:dig-methods} we discuss several different schemes for digitizing $\mathrm{SU}(2)$ as well as the projections to coarser digitizations.
In Section~\ref{sec:num-methods} we describe how we generate $\mathrm{SU}(2)$ lattice ensembles and our classical lattice gauge theory methodology used to compute observables.
In Section~\ref{sec:results}, we present the main results of our work: the error induced by our group-truncated digitization and projection schemes. Finally, in Section \ref{sec:conclusions} we discuss our conclusions.

%===========================================================================8
% Section Separator
%===========================================================================8

\section{Digitizing SU(2)}
\label{sec:dig-methods}

In this work we digitize the gauge group by reducing the infinite set of $\mathrm{SU}(2)$ matrices to a finite set $\mathcal{A}$ of size $|\mathcal{A}|$. 
Each such digitization scheme defines a different surjective mapping $f : \mathrm{SU}(2) \to \mathcal{A}$.
The elements of $\mathcal{A}$ may or may not be elements of $\mathrm{SU}(2)$.
We write $\mathcal{A}_f$ to denote the set $\mathcal{A}$ associated with a given digitization $f$.

There are many different choices for the digitization scheme $f$.
Our objective is to compare several reasonable and well-motivated schemes which reproduce a set of physical observables to a predetermined level of accuracy with the smallest set $\mathcal{A}$, thereby requiring the least resources.
In the following we discuss the different digitization schemes we will study in this work.
We view our list as a base for future improvement rather than an exhaustive study of all possible digitizations.

In what follows, we parametrize any $\mathrm{SU}(2)$ element in the fundamental representation as
\begin{equation}
g = 
\begin{pmatrix}
	a + ib & \;-c + id \\
	c + id & \phantom{\;-}a - ib
\end{pmatrix}
\label{eqn:mx-rep}
\end{equation}
where $a$, $b$, $c$, and $d$ are real numbers in $[-1,1]$. As the determinant of any $\mathrm{SU}(2)$ matrix is unity, $a^2 + b^2 + c^2 + d^2 = 1$ and there are only three independent real degrees of freedom. 

Since $\mathrm{SU}(2)$ and the sphere $S^3$ are diffeomorphic, any $\mathrm{SU}(2)$ matrix may be written as a (Euclidean) four-dimensional unit vector $(a,b,c,d)$.
In this sense, a digitization of $\mathrm{SU}(2)$ is a finite set of points on or near $S^3$.

%===========================================================================8
% Section Separator
%===========================================================================8

\subsection{Fixed-point Digitization}
\label{subsec:fixed}

In typical lattice gauge theory simulations, an $\mathrm{SU}(N)$ matrix is typically represented as $N^2$ complex numbers, with unitarity and the unit determinant condition enforced by hand. Real numbers are typically represented as $64$-bit double-precision floating point numbers (or briefly, doubles).
Hereafter, we refer to this as the ``matrix-of-doubles'' representation of $\mathrm{SU}(N_c)$ matrices.

In a standard double, one bit $s$ represents the sign of the number, $11$ bits represent the integer exponent $-1024 < e < 1023$,\footnote{$2^{11}=2048$.} and the remaining $52$ bits represent the normalized significand.
Denoting the bits of the normalized significand as $m_i$, the significand $1 \le S < 2$ is a fixed-point number whose value is given by
\begin{equation}
S = 1 + \sum_{i=1} \frac{m_i}{2^i}.
\end{equation}
Taken together, the value represented by the $64$ bits of the double is $(-1)^s \times S \times 2^e$. A double can represent values as small as $2^{-1024} \sim 10^{-39}$ and as large as $2^{1024} \sim 10^{39}$, and the values that a double can represent grow exponentially denser closer to zero.
This representation is not optimal. Most glaringly, the real numbers in an $\mathrm{SU}(N)$ matrix are bounded between $-1$ and $1$, so half of the values a double can represent ($e > -1$) are wasted.

A less wasteful representation for the real numbers in an $\mathrm{SU}(N)$ matrix is the simple fixed-point numbers, with the most significant digit starting at $1/2$ \cite{Clark:2009wm}.
Denoting the bits of the fixed-point number as $f_i$, with $f_0=s$ as a sign bit, the value represented by $p$-bits of fixed-point precision is
\begin{equation}
(-1)^s \sum_{i=1}^{p-1} \frac{f_i}{2^i}
\end{equation}
which can represent $2^p$ values evenly spaced between $-1$ and $1$.
This distribution of possible values is better-suited for lattice data: the distribution of values in typical lattice data is closer to flat than to exponentially spiked about zero\footnote{It is straightforward to plot the distribution of $\mathrm{SU}(N)$ values and check that it is closer to flat rather than being exponentially spiked at the origin.}.

Our fixed-point truncation scheme is simply to convert the doubles of an $\mathrm{SU}(2)$ matrix to $p$-bit fixed-point numbers.
More specifically, we first convert the doubles $a$, $b$, and $c$ to 64-bit fixed-point numbers.  
To truncate each number, we cut off the $64-p$ least-significant bits of each number and then, to avoid having to implement fixed-point arithmetic, convert them back to doubles.
To maintain the unit determinant condition to $p$ bits of precision, we compute $d^2=1-a^2-b^2-c^2$ at full precision (while keeping the original sign of $d$) and then apply the same truncation procedure to the resulting value. It is important to note that this operation is only unitary and gauge-invariant to $p$ bits.
We return to this key point in in Section~\ref{sec:results}.

%===========================================================================8
% Section Separator
%===========================================================================8

\subsection{Indexed Mesh Digitization}
\label{subsec:mesh}

The above fixed-point digitization scheme has several drawbacks. 
First, representing an $\mathrm{SU}(N)$ matrix using fixed-point values is still wasteful. 
The independent real degrees of freedom $a$, $b$, $c$ in fixed-point representation parametrize an even grid of $(2^p)^3$ points over the box $(-1,1)^3$, but because $|d| \le 1$ and $d^2 = 1-a^2-b^2-c^2$, all values of $a^2+b^2+c^2 > 1$ are wasted ($\sim 48\%$ of all possible values).
Furthermore, in practice, all of the points lie slightly off the unit sphere and are thus not elements of $\mathrm{SU}(2)$, leading to violations of unitarity and gauge invariance.  In principle, this issue is shared by high-precision floating point numbers.

For a different digitization of the gauge group, we may consider simply choosing some finite subset of $\mathrm{SU}(2)$ elements as the digitization. It is easy to visualize $\mathrm{SU}(2)$ as a unit three-sphere and imagine the subset as a discrete ``mesh'' of $v$ points lying on the sphere. Each element of the subset may then be represented by its index in this subset. This requires only $\text{ceil} \log_2 v$ bits per gauge link\footnote{With $N$ bits we can represent $v=2^N$ numbers. }. 

The most obvious choice for a mesh is a discrete subgroup of $\mathrm{SU}(2)$.
Abelian subgroups are clearly inadequate, since the non-Abelian nature of the group plays a critical role in its nonperturbative dynamics.
Creutz, Jacobs, and Rebbi considered this problem previously from a slightly different perspective~\cite{Creutz:1983ev}.
They found that the finite subgroups of $\mathrm{SU}(2)$ with sufficient non-Abelian structure to avoid significant distortions of physical results are
\begin{enumerate}
\item the $24$-element tetrahedral subgroup $\bar{T}$, 
\item the $48$-element octahedral subgroup $\bar{O}$, and
\item the $120$-element icosahedral subgroup $\bar{I}$ or $\bar{Y}$.
\end{enumerate}
Whether these subgroups are large enough for practical use as a digitization scheme on a quantum computer is an important question, which we revisit in Sec.~\ref{sec:results}.

To get finer digitizations, we can simply pick a larger subset of elements which are distributed approximately evenly across the unit three-sphere. As the three-sphere is generally not diffeomorphically equivalent to the polytope, one cannot find a general exactly uniformly distributed mesh. Geodesic meshes are the familiar solution to this problem in three dimensions and generalize straightforwardly to 4D.

We generate geodesic meshes using the {\tt mvmesh} R package \cite{mvmesh}. To generate the meshes, we use the ``edgewise'' algorithm, which begins with a $4$D octohedron then uniformly tesselates each simplicial face with smaller simplices before ``inflating'' the resulting mesh to an approximate sphere. The package also offers a ``dyadic'' algorithm, which also begins with the octohedron, but recursively tesselates each simplicial face with the simplest simplicial tesselation. These dyadic meshes are more even, but are defined for fewer different values of mesh size $v$. For similar mesh sizes, all observables that we have examined agreed, so we only present results for edgewise meshes.
This approach is similar to the one considered in Ref.~\cite{Lisboa:1982ji}, but less sophisticated in that we do not generate our meshes by subdividing polytopes corresponding to the finite subgroups of $\mathrm{SU}(2)$.

Multiplication in meshes must be implemented using a lookup table \cite{Lisboa:1982ji,Lisboa:1982jj}.  If the mesh is a subgroup of $\mathrm{SU}(2)$, then multiplication is exact. If the mesh is not a closed subset of $\mathrm{SU}(2)$, then the result of multiplying two elements together must be projected back onto the mesh.  In this exploratory study we immediately return to the usual matrix-of-doubles representation after projecting links to meshes, and do not perform any operations in the indexed mesh representation. As such, we are unable to quantify the effect of projection after multiplication.  Our approach is equivalent to using lookup tables to compute traces of multiple mesh elements exactly, as used in Ref.~\cite{Lisboa:1982ji}.

In this study, we project $\mathrm{SU}(2)$ matrices from existing lattice gauge fields computed in the ultrafine digitization to a coarser mesh digitization.
This introduces another potential source of error which compounds the error due to digitization alone, thus it is important to also consider how we perform projections.
To get an idea for how much error is due to projection rather than digitization, we tried several different projection schemes, described below.

%===========================================================================8
% Section Separator
%===========================================================================8

\subsubsection{L2 Norm}

One can project into the mesh by replacing the $\mathrm{SU}(2)$ matrix by its nearest neighbor in the mesh. This requires a metric on the group, for which we use the natural invariant complex matrix normed distance $D(A,B) = ||A-B||$ where $||M|| = \Tr(M^{\dagger} M)$. This amounts to the L2 norm ${\Delta a^2 + \Delta b^2 + \Delta c^2 + \Delta d^2}$ between two points on the three-sphere representation of $\mathrm{SU}(2)$. Consequently, this projection scheme simply chooses the nearest mesh point on $S_3$.

%===========================================================================8
% Section Separator
%===========================================================================8

\subsubsection{Action-Preserving Rounding}

Another idea is to engineer our projection method to preserve physical quantities. 
Ideally, we would like to project each link to the mesh such that all Wilson loops on a lattice are changed as little as possible.
In practice, measuring longer Wilson loops is computationally expensive and finding the exact best projection is  intractable, growing combinatorically with volume $V$ and number of mesh points $v$.
Instead, we define action-preserving rounding (APR) as projection which tries to preserve the local action density.
The Wilson gauge action is a function of the plaquette operator only (as defined in Eq.~(\ref{eqn:plaq})), so this amounts to trying to preserve the value of individual plaquettes.
One may think of this method as trying to project gauge-invariantly. 

In practice, for each plaquette, we replace one link at a time with an element from the mesh, choosing the mesh element which makes the new value of the plaquette closest to the original undigitized value. There is freedom in this algorithm to choose the order in which one replaces the links in each plaquette. Computationally, it is most straightforward to replace all links in each dimension before moving onto another. 
We arbitrarily choose the order $XYZT$ and have not examined the effects of choosing different orderings.

%%%
%Figure
%%%
\begin{figure}[t]
  \centering
\begin{tikzpicture}[scale=0.5, every node/.style={transform shape}]
    \coordinate (Origin)   at (0,0);
    \coordinate (XAxisMin) at (-1.5,0);
    \coordinate (XAxisMax) at (9.5,0);
    \coordinate (YAxisMin) at (0,-1.5);
    \coordinate (YAxisMax) at (0,9.5);

    \draw [line width=0.75mm, black, -latex] (XAxisMin) -- (XAxisMax) node[anchor=north] {\huge $\hat{\mu}$};
    \draw [line width=0.75mm, black, -latex] (YAxisMin) -- (YAxisMax) node[anchor=east] {\huge $\hat{\nu}$};
    
    %Draw the dots 
    \foreach \i in {0,4,8} {
      \foreach \j in {0,4,8} {
        \draw [fill = black](\j,\i) circle (4.0pt);
      }
    }
    \node[label={\Large ${\bm{x}}$}] (a) at (4.0,3.2) {};
    \node[label={\Large ${\bm{x+\hat{\mu}}}$}] (a) at (8.7,3.1) {};

    \begin{scope}[ultra thick, every node/.style={sloped,allow upside down}]
      \draw [line width=0.75mm, red] (4,7.8) -- node{\midarrow}  (4,4.2); 
      \draw [line width=0.75mm, red] (8,4.2) -- node{\midarrow}  (8,7.8); 
      \draw [line width=0.75mm, red] (4.2,4) -- node{\midarrow}  (7.8,4); 
      \draw [line width=0.75mm, red] (7.8,8) -- node{\midarrow}  (4.2,8); 
    \end{scope}

    \node[label={\Large ${\bm{U^{\dagger}_{\nu}(x)}}$}] (a) at (2.8,5.5) {};
    \node[label={\Large ${\bm{U_{\nu}(x + \hat{\mu})}}$}] (c) at (9.8,5.5) {};

    \node[label={\Large ${\bm{U_{\mu}(x)}}$}] (b) at (5.8,2.8) {};
    \node[label={\Large ${\bm{U^{\dagger}_{\mu}(x+\hat{\nu})}}$}] (d) at (6.0,8.0) {};
\end{tikzpicture}
  \caption{The plaquette $\Box_{\mu\nu}(x)$ constructed from the gauge links $U_{\mu}(x)$.}
  \label{fig:IntroPlaquette}
\end{figure}
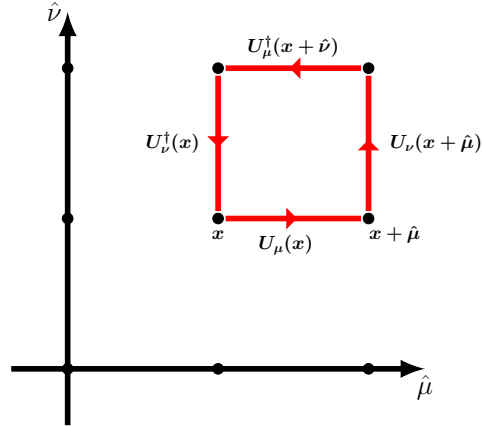

%===========================================================================8
% Section Separator
%===========================================================================8
\section{Classical Lattice Gauge Theory Methodology}
\label{sec:num-methods}

This work uses standard techniques in the classical lattice gauge theory literature; we refer the reader to \cite{Gattringer:2010zz,DeGrand:2006zz} for a discussion of this methodology. For this work, we generate $\mathrm{SU}(2)$ lattice gauge field configurations using MILC code \cite{MILC} adapted to run $N_c=2$. We use the Wilson plaquette action for our discretization of the pure gauge action \cite{Wilson:1974sk}. In each ensemble, we save a gauge configuration after every $1000$ Monte Carlo trajectories, where a trajectory is four overrelaxation steps and one quantum heat bath step. The typical autocorrelation times in our data set are such that $1000$ trajectories is sufficient to decorrelate the observables that we consider.

%%%%%%%%%%%%%%%%%%%%% 
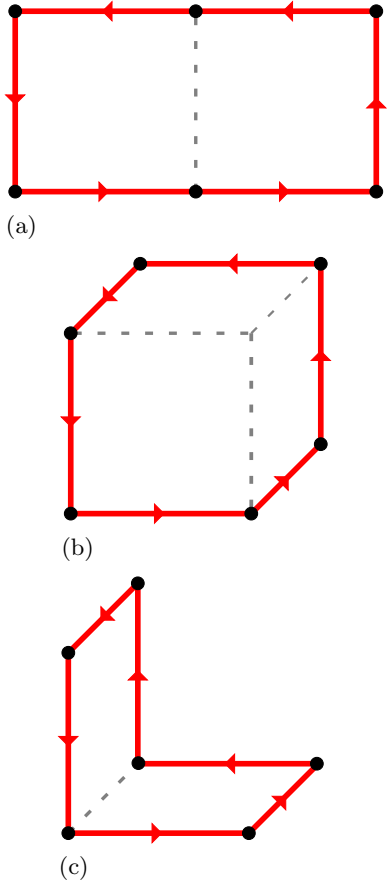
\begin{figure}[t]
  \centering
  \begin{subfigure}[]
    \centering
    \begin{tikzpicture}[scale=0.4, every node/.style={transform shape}]
      \coordinate (O) at (0,0,0);
      \coordinate (A) at (\Depth,0,0);
      \coordinate (B) at (\Depth,\Width,0);
      \coordinate (C) at (0,\Width,0);
      \coordinate (D) at (-\Depth,\Width,0);
      \coordinate (E) at (-\Depth,0,0);

      \begin{scope}[ultra thick, every node/.style={sloped,allow upside down}]
        \draw [line width=0.75mm, red] (O) -- node{\midarrow}  (A); 
        \draw [line width=0.75mm, red] (A) -- node{\midarrow}  (B); 
        \draw [line width=0.75mm, red] (B) -- node{\midarrow}  (C); 
        \draw [line width=0.75mm, red] (C) -- node{\midarrow}  (D); 
        \draw [line width=0.75mm, red] (D) -- node{\midarrow}  (E); 
        \draw [line width=0.75mm, red] (E) -- node{\midarrow}  (O); 
      \end{scope}

      \draw [loosely dashed, line width=0.45mm, gray] (O) --  (C); 
      
      \draw [fill = black] (O) circle (6pt);
      \draw [fill = black] (E) circle (6pt);
      \draw [fill = black] (D) circle (6pt);
      \draw [fill = black] (C) circle (6pt);
      \draw [fill = black] (B) circle (6pt);
      \draw [fill = black] (A) circle (6pt);
    \end{tikzpicture}
  \end{subfigure}
  %%%%%%%%%%%
  \hspace{1cm}
  \begin{subfigure}[]
    \centering
    \begin{tikzpicture}[scale=0.4, every node/.style={transform shape}]
      \coordinate (O) at (0,0,0);
      \coordinate (A) at (0,\Width,0);
      \coordinate (B) at (0,\Width,\Height);
      \coordinate (C) at (0,0,\Height);
      \coordinate (D) at (\Depth,0,0);
      \coordinate (E) at (\Depth,\Width,0);
      \coordinate (F) at (\Depth,\Width,\Height);
      \coordinate (G) at (\Depth,0,\Height);
     
      \begin{scope}[ultra thick, every node/.style={sloped,allow upside down}]
        \draw [line width=0.75mm, red] (O) -- node{\midarrow}  (D); 
        \draw [line width=0.75mm, red] (D) -- node{\midarrow}  (G); 
        \draw [line width=0.75mm, red] (G) -- node{\midarrow}  (F); 
        \draw [line width=0.75mm, red] (F) -- node{\midarrow}  (B); 
        \draw [line width=0.75mm, red] (B) -- node{\midarrow}  (A); 
        \draw [line width=0.75mm, red] (A) -- node{\midarrow}  (O); 
      \end{scope}

      \coordinate (Ai) at ($(A)+(0.10,0   ,0)$);
      \coordinate (Di) at ($(D)+(0   ,0.10,0)$);
      \coordinate (Fi) at ($(F)+(0   ,0   ,0.10)$);
      
      \draw [loosely dashed, line width=0.45mm, gray] (Ai) --  (E); 
      \draw [loosely dashed, line width=0.50mm, gray] (E) --  (Di); 
      \draw [loosely dashed, line width=0.30mm, gray] (E) --  (Fi);

      \draw [fill = black] (O) circle (6pt);
      \draw [fill = black] (D) circle (6pt);
      \draw [fill = black] (G) circle (6pt);
      \draw [fill = black] (F) circle (6pt);
      \draw [fill = black] (B) circle (6pt);
      \draw [fill = black] (A) circle (6pt);
 \end{tikzpicture}
  \end{subfigure}%
  \hspace{1.5cm}
  %%%%%%%%%%%%%%%%%% 
  \begin{subfigure}[]
    \centering
    \begin{tikzpicture}[scale=0.4, every node/.style={transform shape}]
      \coordinate (O) at (0,0,0);
      \coordinate (A) at (0,\Width,0);
      \coordinate (B) at (0,\Width,\Height);
      \coordinate (C) at (0,0,\Height);
      \coordinate (D) at (\Depth,0,0);
      \coordinate (E) at (\Depth,\Width,0);
      \coordinate (F) at (\Depth,\Width,\Height);
      \coordinate (G) at (\Depth,0,\Height);

      \begin{scope}[ultra thick, every node/.style={sloped,allow upside down}]
        \draw [line width=0.75mm, red] (O) -- node{\midarrow}  (D); 
        \draw [line width=0.75mm, red] (D) -- node{\midarrow}  (G); 
        \draw [line width=0.75mm, red] (G) -- node{\midarrow}  (C); 
        \draw [line width=0.75mm, red] (C) -- node{\midarrow}  (B); 
        \draw [line width=0.75mm, red] (B) -- node{\midarrow}  (A); 
        \draw [line width=0.75mm, red] (A) -- node{\midarrow}  (O); 
      \end{scope}

      \coordinate (Ci) at ($(C)-(0.0,0   ,0.05)$);
      \coordinate (Gi) at ($(G)-(0.05  ,0.0,0)$);
      \coordinate (Fi) at ($(F)+(0   ,0   ,0.10)$);
      
      \draw [loosely dashed, line width=0.45mm, gray] (O) --  (C); 
      % \draw [loosely dashed, line width=0.50mm, gray] (E) --  (Di); 
      % \draw [loosely dashed, line width=0.30mm, gray] (E) --  (Fi); 

      \draw [fill = black] (O) circle (6pt);
      \draw [fill = black] (D) circle (6pt);
      \draw [fill = black] (Gi) circle(6pt);
      \draw [fill = black] (Ci) circle(6pt);
      \draw [fill = black] (B) circle (6pt);
      \draw [fill = black] (A) circle (6pt);
   \end{tikzpicture}
  \end{subfigure}
  %%%%%%%%%%%% 
  \caption{The three perimeter six Wilson loops, the (a) rectangle  $\mathcal{P}_1$, the (b) parallelogram $\mathcal{P}_2$, and (c) bent rectangle $\mathcal{P}_3$. Dashed lines are drawn to guide the eye.}
  \label{fig:sketch-loops}
\end{figure}

We generated multiple different ensembles upon which to compute observables. One is a high-statistics zero-temperature ensemble of $1000$ configurations with volume $V=12^4$ at $\beta=2$. The remaining ensembles contain $100$ configurations each. They include 38 finite-temperature ensembles with $V=12^3\times 6$ and additional 6 zero-temperature ensembles with $V=L^4$ for $L \ne 12$. 

For this initial investigation we generate ``undigitized'' ensembles with the standard classical Monte Carlo ultrafine digitization, which uses matrices of $64$-bit doubles to represent gauge links. We do not generate Monte Carlo ensembles with any other coarsely digitized gauge group.

To study the effects of gauge group digitization, we take an undigitized ensemble and project all gauge links to a coarser digitization. For computational convenience, after projecting to the coarser digitization we return the gauge links to the matrix-of-doubles representation. On the resulting digitized ensemble, we measure observables and determine how they have been affected. 

\subsection{Computed Physical Observables}
\label{sec:physics}

This work focuses on the simplest gauge-invariant objects, Wilson loops.
A Wilson loop is the trace of a product of gauge links $U_{x,\mu}$ along a closed loop $\mathcal{L}$
\begin{align}
W_\mathcal{L}[U] = \Tr \prod_{(x,\mu)\in \mathcal{L}} U_{x,\hat{\mu}},
\end{align}
where $x$ denotes the sites in the loop and $\hat{\mu}$ denotes the direction of the link.
Wilson loops are best understood physically through their connection to the potential $V(r)$ between static color charges.
Careful discussions of this connection are available in standard textbooks on lattice gauge theory~\cite{DeGrand:2006zz,Gattringer:2010zz} as well as continuum field theory~\cite{Schwartz:2013pla}.
The present discussion only needs the fact that a rectangular Wilson loop $W$ of spatial length $r$ and temporal length $t$ scales as \cite{Wilson:1974sk}
\begin{align} \label{eq:wloop}
W(t,r) \sim e^{-t V(r)}.
\end{align}

To extract the potential in practice, we construct rectangular Wilson loops for many different values of $t$ and $r$.
At fixed spatial separation $r^\prime$, we fit the lattice data to the decaying exponential in Eq.~(\ref{eq:wloop}) in order to obtain a value for $V(r^\prime)$.
Repeating this process for all spatial separations yields the potential $V(r)$.
These techniques are standard in the lattice literature \cite{DeGrand:2006zz, Gattringer:2010zz}. 
This work however studies the behavior of the potential under different gauge group digitizations.

A Wilson loop of particular importance is the plaquette, the building block of the Wilson plaquette gauge action \cite{Wilson:1974sk}. The plaquette is a $1\times1$ square loop of gauge links, as depicted in Fig.~\ref{fig:IntroPlaquette} and is the simplest gauge-invariant observable that can be measured on a hypercubic lattice.
More quantitatively, the plaquette at site $x$ on the 4D lattice with extent in the $\mu$ and $\nu$ directions is
\begin{equation}
	\Box_{x, \mu\nu} = 
	\Re \Tr \left[
		U_{x, \hat{\mu}}
		U_{x+\mu, \hat{\nu}}
		U_{x+\nu, \hat{\mu}}^\dagger
		U_{x, \hat{\nu}}^\dagger
	\right]
	\label{eqn:plaq}
\end{equation}
where $U_{x, \hat{\mu}}$ is the gauge link in the $\mu$ direction at site $x$.
The three topologically distinct perimeter-six Wilson loops $\mathcal{P}_1$, $\mathcal{P}_2$, and $\mathcal{P}_3$, depicted in Fig.~\ref{fig:sketch-loops}, are longer cousins of the plaquette.
The ensemble expectation values of these quantities are their averages over all orientations, every site on the lattice, and each gauge configuration in the ensemble.
The unimproved Wilson gauge action is a function of the plaquette only, while improved actions are typically functions of longer Wilson loops, like the perimeter-six loops. We include perimeter-six Wilson loops in the study not just as additional observables, but also because we anticipate improved actions may play a role in simulations on quantum computers.

The Polyakov loop is the shortest Wilson loop that winds around the temporal direction of the lattice once, i.e.,
\begin{equation}
	\Omega_{\textbf{x}} = \Tr \prod_{t=1}^{t=N_t} U_{(\textbf{x},t), \hat{t}}
        \label{eqn:polyakov}
\end{equation}
where $U_{(\textbf{x}, t), \hat{\mu}}$ is the gauge link in the $\hat{\mu}$ direction at site $(\textbf{x}, t)$.
In $\mathrm{SU}(N_c)$ pure gauge theory, the Polyakov loop is an order parameter for the finite-temperature deconfinement transition \cite{Svetitsky:1982gs}.
Unlike for $N_c>2$ where it is generally complex, the Polyakov loop is real-valued in $\mathrm{SU}(2)$.
At low temperatures when the system is in a confined phase, the Polyakov loop is protected by a symmetry and therefore vanishes.
Increasing the temperature of the system (corresponding to simulations at larger $\beta$ or shorter $N_t$) eventually results in a phase transition where this symmetry is spontaneously broken and 
the Polyakov loop acquires a nonzero expectation value, which is interpreted as a sign of deconfinement.

%===========================================================================8
% Section Separator
%===========================================================================8

\subsection{Mesh digitization and importance sampling}
\label{sec:mesh_sampling}

Before showing our results, we discuss our expectations for how mesh digitization will affect observables, based on properties of Monte Carlo simulations of lattice field theories.  A naive approach to simulating gauge theories would generate $\mathrm{SU}(N)$ gauge field configurations with gauge links randomly distributed by the Haar measure, and weight them by the action term $e^{-S}$ when computing observables\footnote{This approach is intractably inefficient, as the vast majority of possible gauge configurations are exponentially suppressed by the action term.}. Monte Carlo simulations instead use importance sampling, which includes the $e^{-S}$ term as part of the measure when randomly generating configurations. This induces correlations (over an ensemble of gauge configurations generated with importance sampling) between all gauge links.

The Haar measure has the property that
\begin{equation}
  \int dU \Tr U = 0
  \label{eqn:HaarProp}
\end{equation}
where $U$ is an $\mathrm{SU}(N)$ matrix.
The integral Eq.~(\ref{eqn:HaarProp}) can be thought of as the expectation value of a Wilson loop in a theory where $e^{-S} = 1$.
It follows that the correlations between gauge links due to the $e^{-S}$ term are what allow for nonzero Wilson loops, and that the expectation value of any Wilson loop constructed from completely random $\mathrm{SU}(N)$ matrices is zero.
We thus expect that making gauge links more random (or equivalently, less correlated) will suppress the expectation values of Wilson loops.

To interpret the systematic error seen in our digitized results below in Sec.~\ref{sec:results}, note that projecting an ultrafine-digitized $\mathrm{SU}(2)$ matrix to a coarser digitization can be thought of as displacing that matrix in the group manifold.
Displacements applied to different gauge links are completely uncorrelated for fixed-point truncation and projection to meshes with the L2 scheme, and less so for APR projection.
We can thus think of the effect of truncation or projection as random kicks which add incoherent noise to the gauge links.
This noise washes out correlations between gauge links induced by importance sampling.

\begin{figure}[t]
  \centering
  \includegraphics[width=0.49\textwidth]{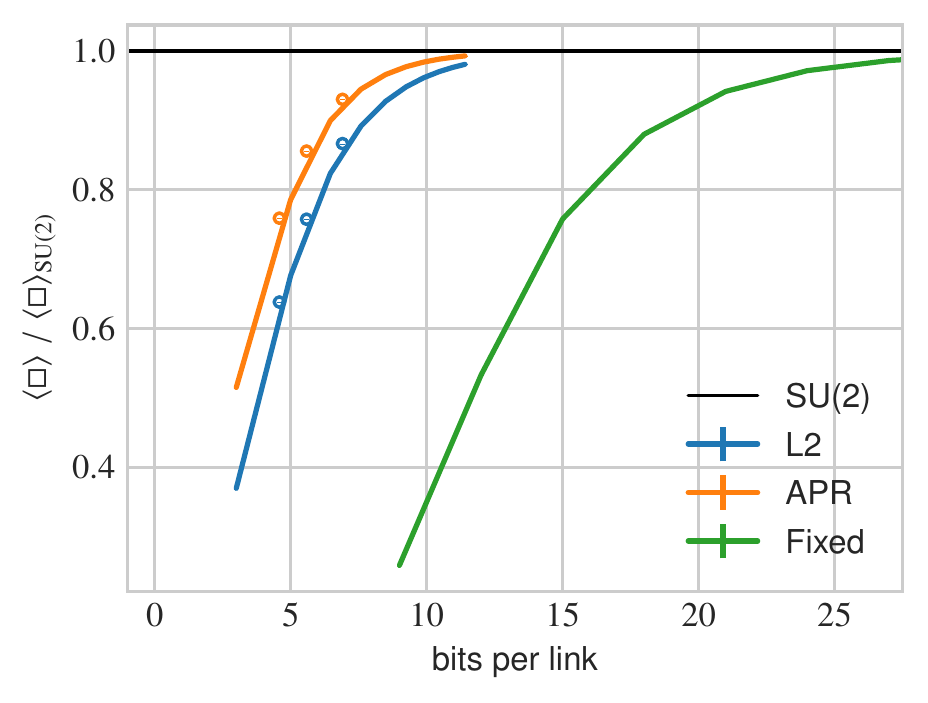}
  \caption{
    The relative error in the plaquette expectation value (defined in Sec.~\ref{sec:physics})  as a function of bits-per-link for different digitization and projection schemes, normalized by its undigitized value. These results are computed on an ensemble of $1000$ configurations with $V=12^4$ and $\beta=2$. Representing an $\mathrm{SU}(2)$ link as a mesh element requires $\log_2 v$ bits, where $v$ is the number of mesh points. A fixed-point digitization of precision $p$ requires $3p$ bits. In the legend, $\mathrm{L}2$ labels projection to a mesh using the $\mathrm{L}2$ norm, while APR labels projection using the action (plaquette) preserving scheme, both defined in Sec.~\ref{subsec:mesh}. Circles indicate projections onto the finite subgroups $\bar{T}_{24}$, $\bar{O}_{48}$, and $\bar{I}_{120}$. }
  \label{fig:plaq-vs-bc}
\end{figure}

Taken together, this suggests that projection to coarser meshes will disrupt the correlations between gauge links induced by importance sampling and thereby suppress the expectation values of Wilson loops.
When projecting to finer meshes the correlations between gauge links are damped but remain significant. However, when projecting to increasingly coarser meshes correlations become small and the gauge links appear random within $\mathrm{SU}(2)$. In this case, it becomes comparable to performing a path integral with $e^{-S}=1$ and restricting the group integration to those matrices in the mesh, approximating the integral over $\mathrm{SU}(2)$ in Eq.~(\ref{eqn:HaarProp}).

As discussed in Sec.~\ref{sec:results}, our results are broadly consistent with this narrative.
The values of all Wilson loops that we measure are increasingly suppressed by projections to coarser digitizations.
Furthermore, correlations over longer distance scales (i.e., large scale structure in the gauge fields) should be more robust against the addition of incoherent noise compared to correlations over short distance scales (i.e., fine structure in the gauge fields). We observe that for a particular digitization and projection scheme the error induced in the static potential $V(r)$ is less than that induced in any of the shorter-range observables that we examine.  We also see the error in the static potential $V(r)$ smoothly decreases as $r$ increases (cf.~Fig.~\ref{fig:pot_mesh_vs_r} below).

%===========================================================================8
% Section Separator
%===========================================================================8

\section{Results}
\label{sec:results}

Here we empirically quantify the systematic error that is introduced by the different digitization and projection schemes that we have studied.  We compare the different schemes based on the number of bits required to represent a gauge link, or ``bits-per-link''. If $p$ is the number of bits of precision for each fixed-point number (including the sign bit), and it takes three fixed-point numbers to represent an $\mathrm{SU}(2)$ matrix parsimoniously, then the bits-per-link is $3p$ for a fixed-point digitization. To be explicit, in this scheme there can be $2^{3p}$ (potentially bad\footnote{See Sec.~\ref{subsec:fixed} about the wastefulness of the fixed-point representation.}) different representations of $\mathrm{SU}(2)$ matrices to $p$ bits of precision. On the other hand, for a mesh of size $v$ the number of bits-per-link is the number of bits required to index the mesh, $\log_2 v$. Again, to give an explicit example, a mesh with $v=4$ $\mathrm{SU}(2)$ matrices requires a two-bit index to provide $2^2=4$ unique labels, with one label for each matrix. In practice we do not have fractional bits, so $\log_2 v$ must be rounded up to the nearest integer. We do not do this here to keep our curves smooth so that the reader may interpolate.

\begin{figure*}[t]
  \centering
  \includegraphics[width=0.99\textwidth]{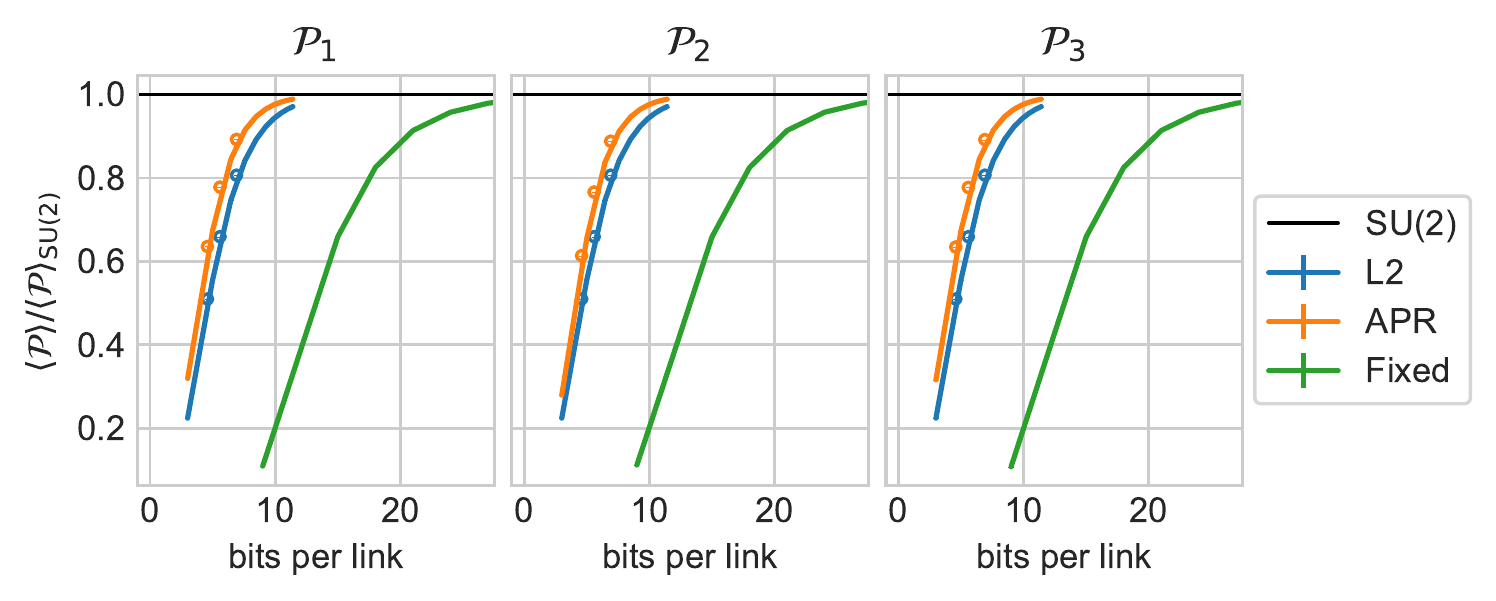}
  \caption{
    The relative systematic error from digitizing the three topologically-distinct perimeter-six Wilson loops (shown in Fig.~\ref{fig:sketch-loops}) as a function of bits-per-link for different digitization schemes, normalized to its undigitized value. These results are computed on an ensemble of $1000$ configurations with $V=12^4$ and $\beta=2$. Refer to Fig.~\ref{fig:plaq-vs-bc} for more details.	}
  \label{fig:six-vs-bc}
\end{figure*}

We now show our results. First, in Fig.~\ref{fig:plaq-vs-bc}, we show the plaquette expectation value against bits-per-link for different digitization and projection schemes. We empirically find that decreasing the bits-per-link of our digitization schemes reduces the plaquette expectation value towards zero, consistent with our arguments in Sec.~\ref{sec:mesh_sampling}. The fixed-point scheme performs drastically worse than the mesh-based schemes, requiring at least twice as many bits to achieve the same error.  In part, this is due to the wastefulness of the fixed-point representation as discussed in Sec.~\ref{subsec:fixed}.  Additionally, each matrix is only unitary to $p$-bit precision in this scheme, in contrast to the mesh-based schemes where each matrix is exactly in $\mathrm{SU}(2)$.
For the mesh-based schemes, projection with APR outperforms projection using the L2 norm, especially on finer meshes, where the induced systematic error is less than half the magnitude.
We observe this discrepancy throughout our data, which suggests that the dominant source of error in our data is projection, rather than inherent to digitization.
Projections onto discrete subgroups induce slightly less error than projections onto  geodesic meshes of equivalent size.  However, projections to sufficiently fine geodesic meshes outperform even the largest discrete subgroup of $\mathrm{SU}(2)$.

\begin{figure*}[t]
  \centering
  \includegraphics[width=0.99\textwidth]{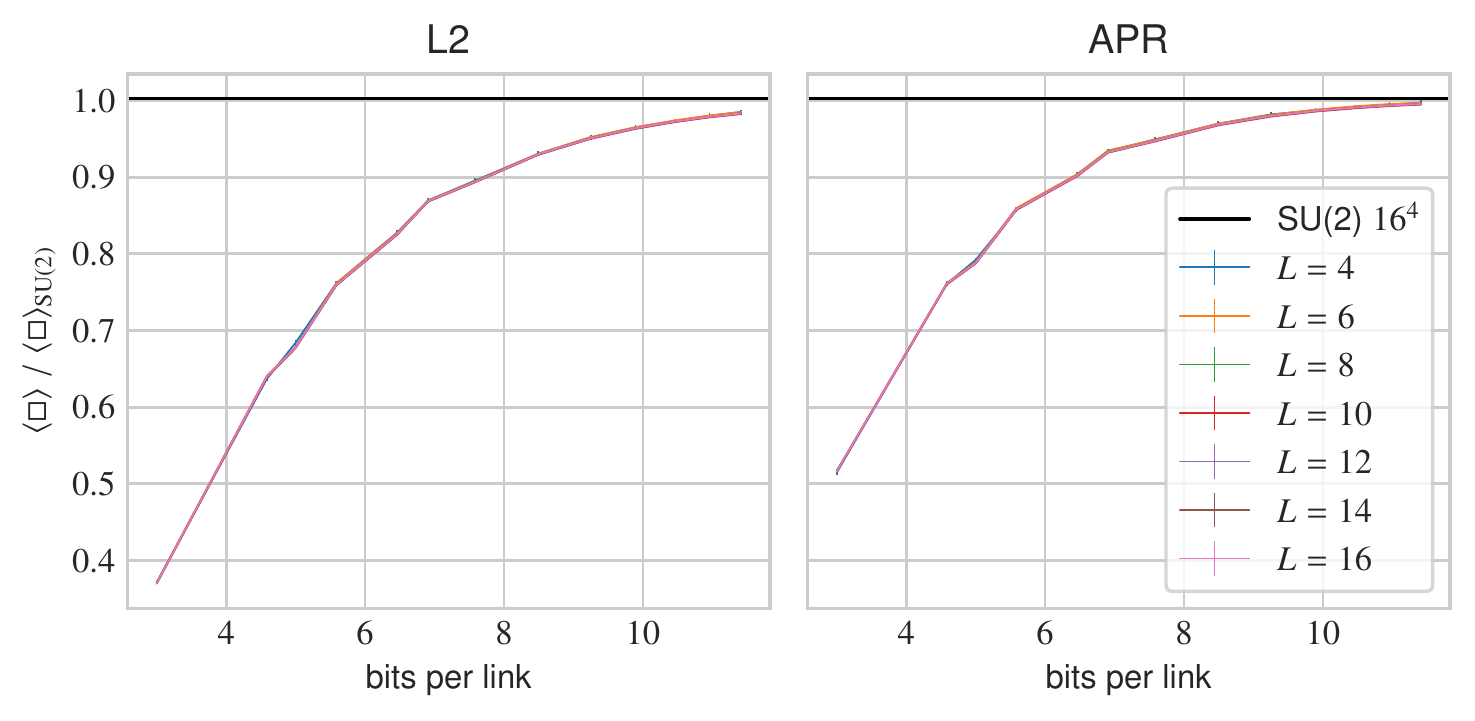}
  \caption{
    The plaquette expectation value (defined in Sec.~\ref{sec:physics}) as a function of bits-per-link for different lattice volumes, normalized to the undigitized value for $16^4$.
    These results are computed on ensembles of 100 configurations with $V=L^4$ and $\beta=2$ (except for the $V=12^4$, which has 1000 configurations).
    Refer to Fig.~\ref{fig:plaq-vs-bc} for more details.
  }
  \label{fig:plaq-finite-volume}
\end{figure*}

Next, Fig.~\ref{fig:six-vs-bc} shows the effects of projection and digitization on the expectation values of the three perimeter-six Wilson loops.
These observables are affected by digitization similarly to the plaquette (cf. Fig~\ref{fig:plaq-vs-bc}).
It also appears that each operator at some specific bits-per-link is suppressed by the same factor consistent with the hypothesis of projection adding incoherent noise to the gauge links, as discussed in Sec.~\ref{sec:mesh_sampling}. 

Fig.~\ref{fig:plaq-finite-volume} shows the volume dependence of the effect of projection and digitization on the plaquette expectation value. We examine $L^4$ lattices varying $L$ while keeping all other physical scales (i.e., $\beta$) fixed. The $L=12$ data of this plot is the same as is used in Fig.~\ref{fig:plaq-vs-bc}.
As can be observed, curves for different volumes essentially overlap on this $y$-axis scale, indicating that the volume dependence of the error is much smaller than the mesh size dependence (for the range plotted). We see similar volume independence in the three perimeter-six Wilson loops. Although we have not seen any significant volume dependence, this result may be observable-specific as the plaquette and perimeter-six Wilson loops are short distance quantities which only require a small volume in order to saturate.

\begin{figure*}[t]
  \centering
  \includegraphics[width=0.99\textwidth]{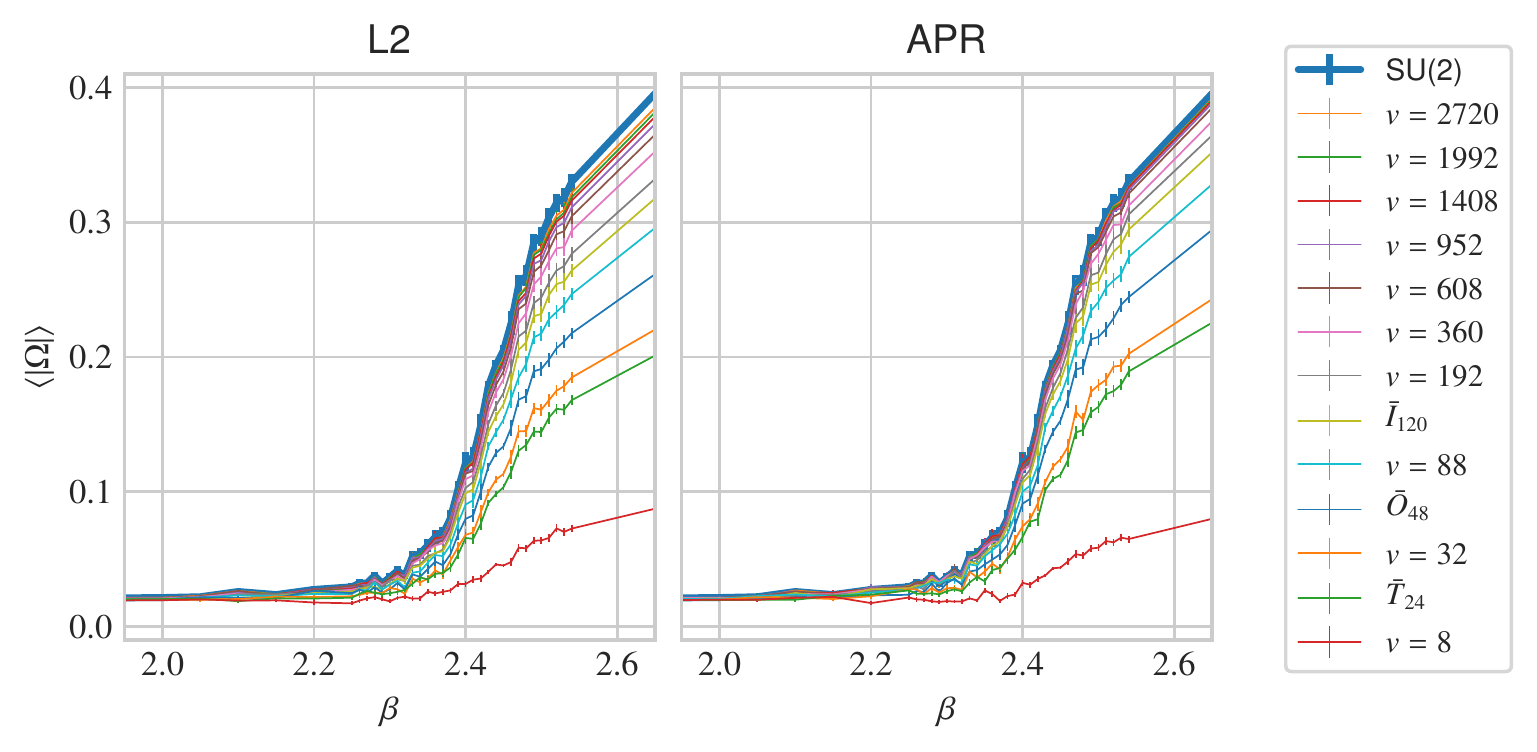}
  \caption{
    The expectation value of the absolute value of the Polyakov loop (defined in Eqn.~(\ref{eqn:polyakov})) as a function of {bare gauge coupling} $\beta=2N/g^2$ for different mesh-based digitization schemes. These results are computed on ensembles of $100$ configurations with $V=12^3\times 6$ and with $\beta$ values interpolated across the deconfinement transition. The data points are indicated by the presence of error bars, and interpolated between for the readers convenience. Refer to Fig.~\ref{fig:plaq-vs-bc} for more details. 
  }
  \label{fig:meshed-pl-vs-beta}
\end{figure*}

\begin{figure*}[t]
  \centering
  \includegraphics[width=0.99\textwidth]{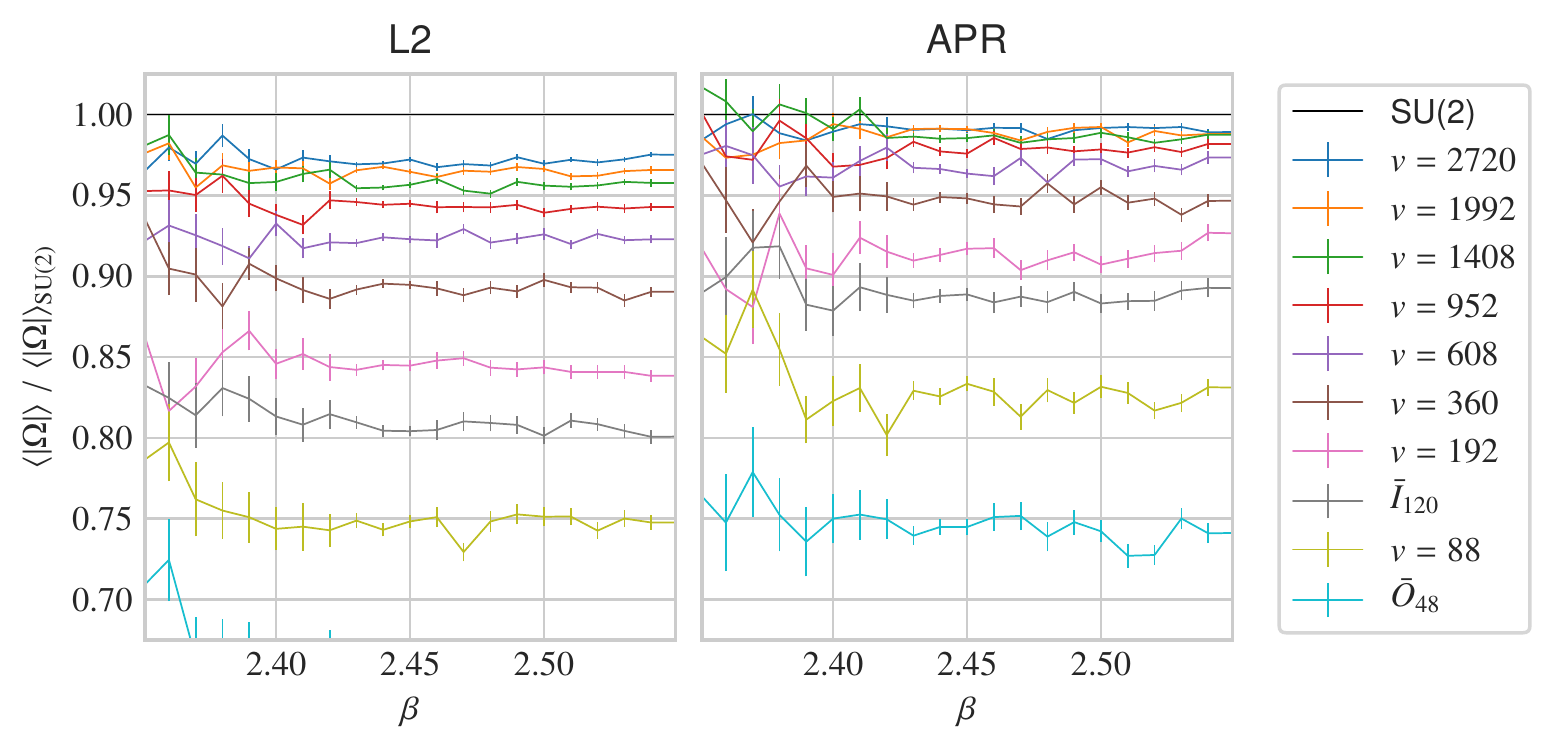}
  \caption{
    The relative systematic error from digitizing the absolute value of the Polyakov loop as a function of $\beta=2N/g^2$ for different mesh-based digitization schemes. Note that the denominator of the y-axis is the undigitized value for each ensemble.  This data is the same as in Fig.~\ref{fig:meshed-pl-vs-beta}. Refer to Fig.~\ref{fig:meshed-pl-vs-beta} for further details.
  }
  \label{fig:meshed-pl-error-vs-beta}
\end{figure*}

In Fig.~\ref{fig:meshed-pl-vs-beta}, we examine how different mesh digitizations affect the Polyakov loop expectation value as a function of the bare gauge coupling $\beta=2N/g^2$.
As with the other observables, digitization and projection suppresses the Polyakov loop expectation value at all $\beta$ values.
Figure~\ref{fig:meshed-pl-error-vs-beta} plots the relative systematic error induced by the different digitization schemes for the data shown in Fig.~\ref{fig:meshed-pl-vs-beta}.  
The Polyakov loop is close to zero in the confined phase, and so we see predominantly noise at lower $\beta$s.
However, in the deconfined phase, the curves in Fig.~\ref{fig:meshed-pl-error-vs-beta} appear to be flat, indicating that the effect of the digitization for each $\beta$ value is simply an overall multiplication by a constant smaller than one.
Figure~\ref{fig:meshed-pl-error-vs-bits} shows the relative error averaged over the range $2.4 \leq \beta \leq 2.6$ as a function of bits-per-link, making it clear that this multiplicative constant approaches zero as the bits-per-link are reduced, again consistent with our arguments in Sec.~\ref{sec:mesh_sampling}.
Figure~\ref{fig:meshed-pl-error-vs-bits} also shows convergence to the undigitized result explicitly.
Projection with APR produces less error than with the L2 norm and appears to converge to the undigitized value quicker, but our data are unable to determine whether any systematic error survives in the limit of large bits-per-link for either scheme.
We see no error due to digitization and projection in the critical value of $\beta$ where the system deconfines, a positive indication as projection should not change the phase dynamics.

\begin{figure}
  \centering
  \includegraphics[width=0.49\textwidth]{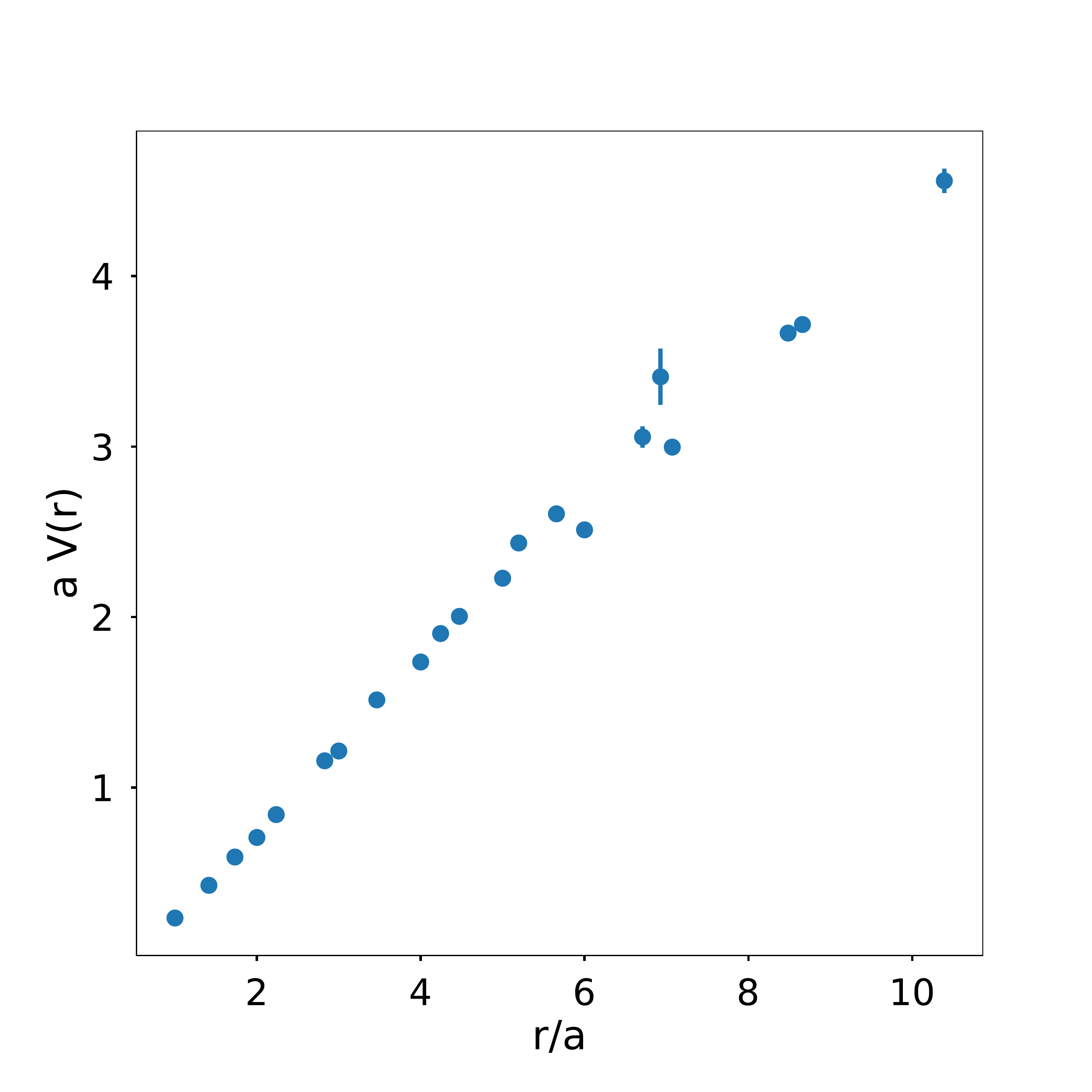}
  \caption{
    The lattice static potential (defined in Eq.~(\ref{eq:wloop})) as a function of distance in lattice units. The data in this plot was generated using an ensemble of 1000 configurations with $V=12^4$ and $\beta=2.0$. We refer readers to the text for further details.
  }
  \label{fig:potential}
\end{figure}

Finally, we turn to the static potential. Fig.~\ref{fig:potential} shows the static potential $aV(r)$ as a function of distance $r/a$, computed in the usual lattice QCD ultrafine digitization.
Due to the large lattice spacing of this ensemble (i.e., the strong bare coupling), the potential is dominantly linear in all distance scales in our simulation.
Above $r/a \approx 6$, the data become unreliable due to the exponentially decreasing signal in the Wilson loop as shown in Eq.~(\ref{eq:wloop}). We restrict our subsequent discussion and figures to the region $r/a \lesssim 6$.

\begin{figure*}[t]
  \centering
  \includegraphics[width=0.99\textwidth]{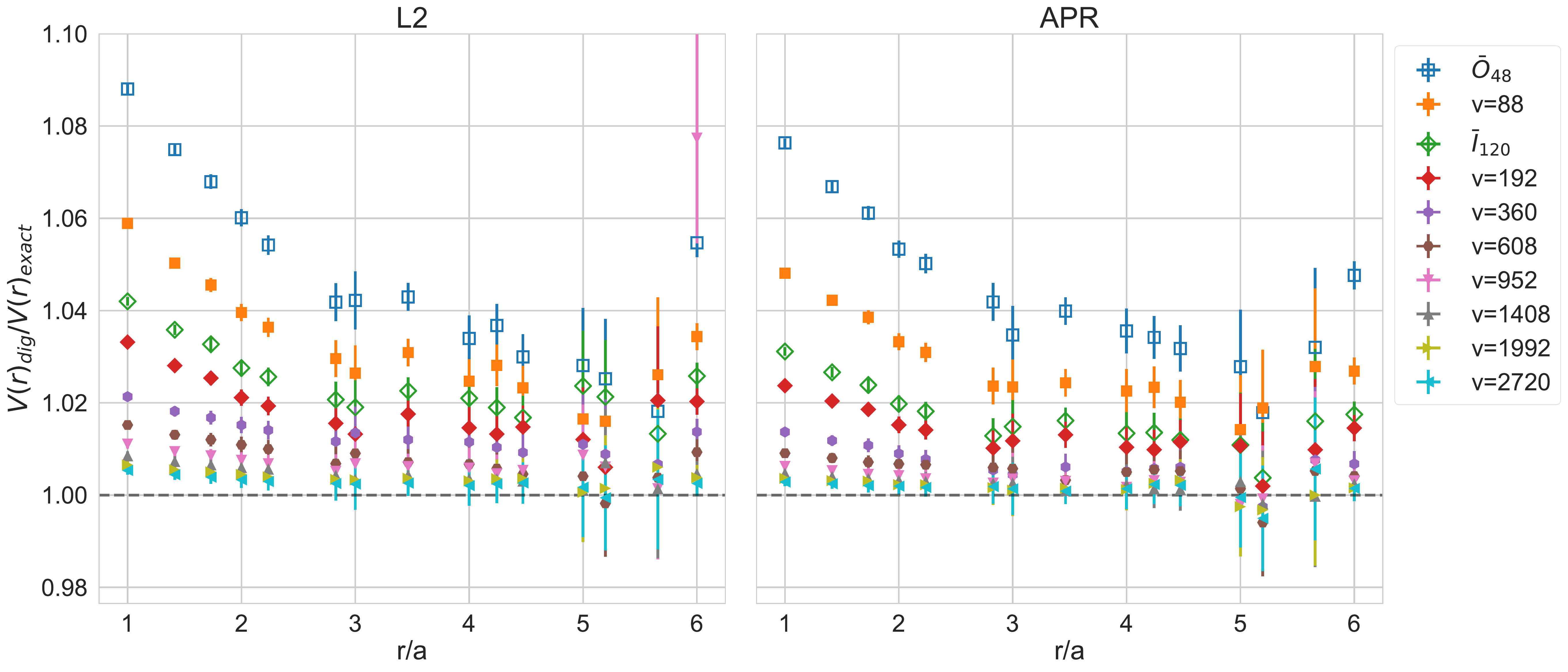}
  \caption{
    The digitization error in the static potential $V(r)$ as a function of distance.
    Filled symbols denote edgewise meshes, while open symbols denote discrete subgroups of SU(2).
    The left pane shows the L2 projection, while the right pane shows the APR projection.
  }
  \label{fig:pot_mesh_vs_r}
\end{figure*}

Fig.~\ref{fig:pot_mesh_vs_r} shows the digitization error in the static potential as a function of distance.
The most interesting feature of this figure is the distance dependence. For any given mesh size, within our statistical precision, the error induced by digitizing $V(r)$ decreases with distance until saturating around $r/a \approx 3$ where our statistical error becomes appreciable. It is also worth noting that the static potential gets larger as the bits-per-link gets smaller, a consequence of the expectation values of Wilson loops approaching zero for projections to coarser digitizations.
The APR projection outperforms the $\mathrm{L}2$ projection at short distances.
At longer distances $r/a \gtrsim 3$ the situation is not as clear: APR appears to perform slightly better, but this is not statistically significant for finer meshes.

Fig.~\ref{fig:pot_error_vs_size} gives the digitization error in the static potential as a function of bits-per-link. There are several interesting features of this data. For a fixed number of bits-per-link, the systematic error from digitizing the static potential is much smaller than the systematic error in the plaquette, Wilson loops, or Polyakov loop. For example, with six bits-per-link the systematic error in the static potential at $r/a=2.0$ can be as low as $6\%$, while the other quantities have at least $15\%$ error.
As expected, the digitized potential converges to the usual lattice QCD ultrafine-digitized result as the mesh becomes sufficiently large, with as few as $10$ bits-per-link being indistinguishable from the ultrafine digitization. As expected from Fig.~\ref{fig:pot_mesh_vs_r}, we observe less systematic error at larger distances.

Figures~\ref{fig:pot_mesh_vs_r} and~\ref{fig:pot_error_vs_size} convey that long-distance physics is, evidently, less sensitive to digitization.
In Sec.~\ref{sec:mesh_sampling} we argue that projecting to a coarse digitization is roughly akin to adding uncorrelated random noise to all the gauge links.
The large effect at short length scales, approaching the Coulombic region of the potential, is consistent with what we have already observed in the plaquette and perimeter-six Wilson loops.
At longer distances, we see a much smaller effect.
It appears that correlations over longer length scales are less susceptible to the addition of incoherent noise, consistent with our arguments in Sec.~\ref{sec:mesh_sampling}.
The statistically insignificant difference between APR and L2 projections at long distances suggests that projection is no longer the dominant source of error at long distances.
If this is the case, then the error purely due to digitization is already at the sub-percent level at 9 bits-per-link.

%===========================================================================8
% Section Seperator
%===========================================================================8

\section{Discussion and Conclusions}
\label{sec:conclusions}

\begin{figure}
  \centering
  \includegraphics[width=0.49\textwidth]{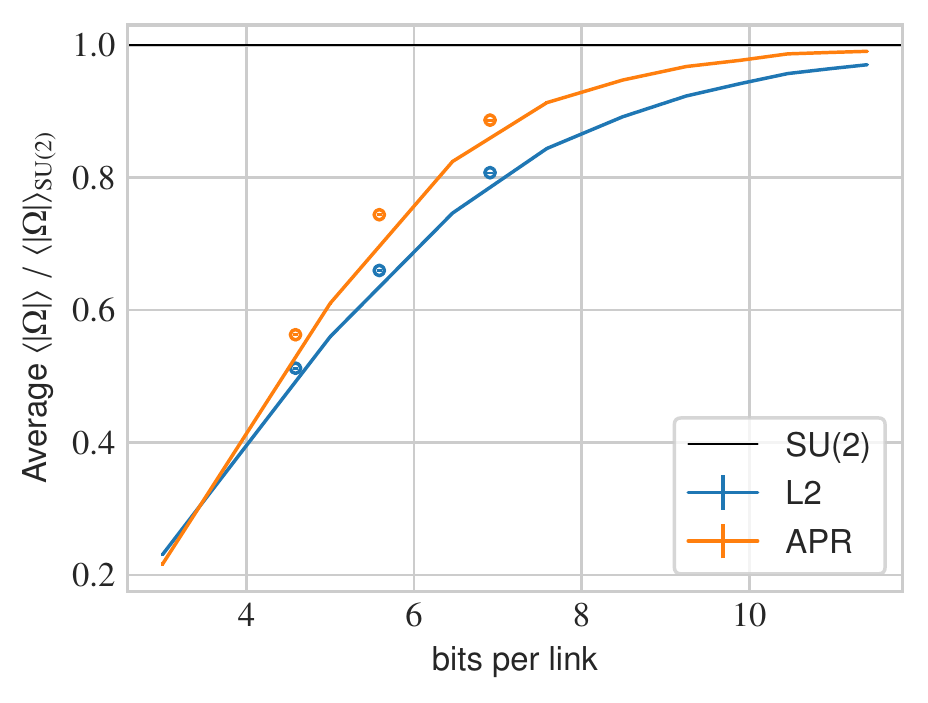}
  \caption{
    The relative systematic error from digitizing the absolute value of the Polyakov loop as plotted in Fig.~\ref{fig:meshed-pl-error-vs-beta}, here averaged over $\beta$ in the range from $2.4$ to $2.6$, as a function of bits-per-link.  The curves are results for geodesic meshes, while points are results for discrete subgroups.
  }
  \label{fig:meshed-pl-error-vs-bits}
\end{figure}

\begin{figure*}[t]
  \centering
  \includegraphics[width=0.99\textwidth]{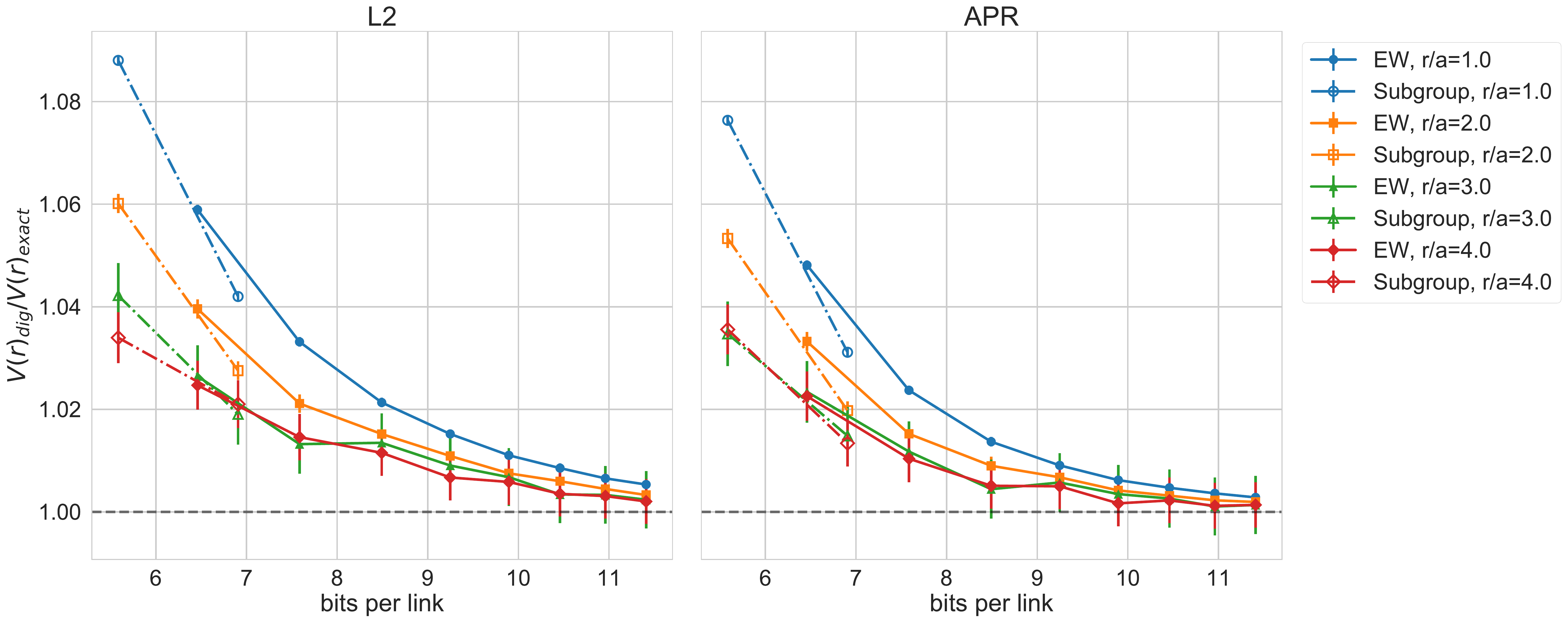}
  \caption{
    The relative error from digitizing the static potential (defined in Eqn.~(\ref{eq:wloop})) as a function of bits-per-link.           Each line in this figure corresponds to a different fixed value of the distance $r/a$.
    We refer the reader to Fig.~\ref{fig:potential} for further details on the ensemble data used to generate this plot.
    Fig.~\ref{fig:plaq-vs-bc} contains additional details on terminology.
  }
  \label{fig:pot_error_vs_size}
\end{figure*}

In this work we have empirically quantified the systematic error associated with digitizing the $\mathrm{SU}(2)$ gauge group in different ways using classical lattice gauge theory. Section \ref{sec:results} contains the main results of this work. Figs.~\ref{fig:plaq-vs-bc}, \ref{fig:six-vs-bc}, \ref{fig:meshed-pl-error-vs-bits}, and \ref{fig:pot_error_vs_size} show the relative systematic error in the plaquette, the perimeter-six Wilson loops, the Polyakov loop, and the static potential as a function of the (qu)bit requirements, for each digitization and projection scheme, in units of bits-per-link. Across all observables, we observe several behaviors consistently. Projection to a coarser digitization suppresses the values of Wilson loops, consistent with our arguments in Section \ref{sec:mesh_sampling}.	 For a given mesh, action-preserving rounding induces less error and asymptotes more quickly to the undigitized result than L2 norm projection.
Projecting to a subgroup mesh appears to induce less error than projecting to a geodesic mesh of similar size, but the finer geodesic meshes outperform the largest discrete subgroup of $\mathrm{SU}(2)$.
Finally, we also observe that long-distance physics is less sensitive to digitization and projection than short-distance physics.

Taken together, our results indicate that O(10) bits per link suffice to capture the essential physics of SU(2) gauge theory.
Compared to modern classical lattice simulations, which default to $512$ bits-per-link using double-precision floating-point numbers, this is an improvement of nearly two orders of magnitude. This observation has important implications for the types of physics that will be accessible to NISQ era quantum computers.

Moreover, the digitization and projection schemes discussed above amounts to forms of lossy compression for gauge links. The digitization error shown in this work is smaller than other dominant sources of error in many current classical lattice gauge theory calculations, and mesh-based schemes offer more than an order of magnitude compression over the commonly-used floating point representations. Thus, this work may have applications in modern classical lattice calculations if link compression can be used to overcome bandwidth bottlenecks.

We found that mesh-based schemes dramatically outperform the truncated fixed-point scheme, achieving similar accuracies with less than half the bits-per-link.
Some of this may be attributed to the wastefulness of the fixed-point representation as discussed in Sec.~\ref{subsec:mesh}, but this only accounts for a single bit of the difference.
More importantly, each fixed-point representation matrix is only unitary to $p$ bits of precision, and unitarity is a key building block of gauge theories.
This emphasizes an important lesson: different digitization schemes are possible, but standard principles of quantum physics remain a guiding light for constructing optimal digitizations as new technology is explored.\footnote{It was also necessary to enforce gauge invariance and locality when applying machine learning techniques to lattice QCD in order to produce results \cite{Shanahan:2018vcv}.}

If gauge links are represented in an indexed mesh representation, multiplication must be implemented using lookup tables. There is existing work which considers how to do this efficiently on classical computers \cite{Lisboa:1982ji,Lisboa:1982jj}. An important question for future research is whether this can be easily and efficiently implemented on a quantum computer.

In this study, we examined SU(2) pure gauge theory.
However, we are obviously more interested in QCD, whose gauge group is SU(3).
Repeating this study for SU(3) pure gauge theory and for theories with dynamical fermions are obvious next steps.
We note that previous work which used indexed mesh digitizations to simulate SU(3) pure gauge theory found that a mesh with 1080 elements was too coarse to avoid artifact phases near interesting values of $\beta$, but a mesh with 38880 elements was viable \cite{Bhanot:1981xp}.
Translating these numbers to our bits-per-link metric, this suggests that $\sim 10$ bits is insufficient to represent an SU(3) gauge link, but $\sim 15$ bits may be enough.

As noted above, we do not generate coarsely digitized ensembles. Systematically studying the error induced when simulating using a coarse digitization is an interesting and complementary direction for future work.
In this study, we generate ensembles in the standard lattice ultrafine digitization, then project to coarser digitizations.
This introduces error specific to projection, which is difficult to disambiguate from error due to digitization alone.
Generating data directly in a coarse digitization does not require projection, and thus would allow estimation of digitization error without this confounding factor.
However, simulating with coarse digitizations would introduce new errors that our method is immune to, and it may be that these errors are large (cf. the lattice artifact phases seen by Refs.~\cite{Rebbi:1979sg,Petcher:1980cq,Bhanot:1981xp}). Thus, coarsely digitized simulation and our approach of projection can provide independent probes of digitization error.

Our results are already very promising, but it is likely possible to reduce (qu)bit requirements even further.
One possible method is by using better meshes and projection schemes.
We observed that projecting to discrete subgroups of $\mathrm{SU}(2)$ induced less error than projecting to geodesic meshes.
This suggests that meshes made by interpolating between points in discrete subgroups (as proposed in Ref.~\cite{Lisboa:1982ji}) might perform better than geodesic meshes.
Our action-preserving rounding method for projecting onto meshes performs substantially better than projection using the L2 norm, even on fine meshes where one might expect the projection method to matter less.
Pushing further and developing more sophisticated action-preserving or gauge-invariant projection schemes would be an interesting direction for future work.
This is a particularly interesting direction for further thought when considered in the context of using classical lattice simulations to prepare states for quantum computers, and for gauge link compression for high-performance computing.

Another way of reducing (qu)bit resource requirements is by removing non-physical degrees of freedom from the theory.
The authors of \cite{Kaplan:2018vnj} show that when using an eigenstate truncation method the vast majority of states in the naive finite-dimensional Hilbert space are unphysical. There they explore how to construct the theory on the physical subspace alone. It would be worthwhile to explore if such techniques could also be applied to mesh digitizations. Similarly, implementing a digitization scheme which uses gauge-fixing could also reduce (qu)bit resource requirements. Exploring the mapping between the eigenstate truncation method described in Refs.~\cite{Klco:2018kyo,Lu:2018pjk,Zohar:2013zla} and the group-value digitization scheme explored in this work may provide further insights.

Our work is particularly applicable to the formulation of $\mathrm{SU}(N)$ gauge theories (with matter fields) on quantum computers \cite{Brower:1997ha, Zohar:2014qma}. In these formulations, the $\mathrm{SU}(N)$ link matrices build the quantum-link Hamiltonian. Our results for  four-dimensional pure gauge $\mathrm{SU}(2)$ indicate that only a small number of (qu)bits per link may be needed to achieve accurate results in these theories. If QCD is formulated on a quantum computer in this way, future work is needed to empirically quantify the (qu)bit requirements of these $\mathrm{SU}(N)$ implementations.

%===========================================================================8
% Section Separator
%===========================================================================8
\section*{ACKNOWLEDGMENTS}

We would like to thank Tom DeGrand and Andreas Kronfeld for useful discussions. 
We are also grateful to the MILC collaboration for the use of the source code adapted to generate the ensembles in this study.  This work was supported in part by the U.~S.~Department of Energy under contract DE-SC0010005 (DH and EN).  This manuscript has been authored by Fermi Research Alliance, LLC under Contract No.~DE-AC02-07CH11359 with the U.~S.~Department of Energy, Office of Science, Office of High Energy Physics. Brookhaven National Laboratory is supported by the U.~S.~Department of Energy under contract no.~DE-SC0012704.  CH and KH would like to thank the University of Colorado Boulder for hosting each, which directly led to this work.

%===========================================================================8
% Section Separator
%===========================================================================8
\bibliography{LatticeDigitization}

\end{document}